\documentclass[twoside]{article}
\usepackage{fleqn,espcrc2}
\usepackage{graphicx}
\title{Low-Energy Hadronic Interaction Models}

\author{D.~Heck\address{Forschungszentrum Karlsruhe, 
                        Institut f{\"u}r Kernphysik, 
                        P.O. Box 3640, D--76021 Karlsruhe,
                        Germany}\thanks{e-mail: dieter.heck@ik.fzk.de}
       }
\begin{document}

\begin{abstract}

Different interaction models treating hadronic collisions below
$E_{\rm lab}$$<$$100$ GeV in air shower simulations are compared within
the frame of the CORSIKA program.  
Their behaviour is studied in isolated collisions of protons or pions 
with light-mass targets in comparison with experimental data.  
It is discussed how model properties influence air-shower 
parameters like the longitudinal profile or the lateral distribution 
of particles arriving at ground. 
\vspace{1pc}
\end{abstract}

\maketitle

\section{INTRODUCTION}

In extensive air showers (EAS) induced by cosmic hadronic particles
low-energy secondary hadrons collide with the atmospheric nuclei 
thus forming the final branches of the hadronic shower backbone.
Recently Ref.~\cite{drescher} reported that in simulations the lateral
particle distribution at core distances $>$1 km shows a strong 
dependence on the used low-energy hadronic interaction model.
Therefore several codes describing low-energy 
interactions have been examined with the EAS simulation program 
CORSIKA \cite{corsika} in the present study. 
Preliminary results are given in \cite{heck_tsukuba}. 
The used interaction codes are presented in Sect.~\ref{sec-models}.
In Sect.~\ref{sec-sing_coll} their predictions are 
compared with experimental results of single hadronic collisions
with light targets close in 
mass to the atmospheric constituents.
How the behaviour in single collisions transforms into measurable 
properties of EAS is discussed in Sect.~\ref{sec-shower_sim}.
The consequences to simulations are exemplified in
Sect.~\ref{sec-implicat}.

\section{MODELS}
\label{sec-models}

Within this study the 4 codes 
FLUKA \cite{fluka}, GHEISHA \cite{gheisha}, Hillas Splitting
Algorithm \cite{hillas}, and UrQMD \cite{urqmd}  are 
investigated in view of their application 
to low-energy ($E_{\rm lab}$$<$100 GeV) hadronic interactions
within the EAS simulation program CORSIKA.

In some comparisons the high-energy interaction models DPMJET II.55
\cite{dpmjet}, {\sc neXus 3} \cite{nexus}, QGSJET$~$01 \cite{qgs},
or SIBYLL$~$2.1 \cite{sibyllnew} are included overstretching their
recommended energy range to show up possible problems in the 
transition region around 80 GeV.

The hadronic event generator FLUKA 2002 \cite{fluka} is used with CORSIKA
only for the description of the inelastic interactions with
laboratory energies below several 100 GeV. 
Within FLUKA these collisions are handled by different 
hadronic interaction models above, around, and below the 
nuclear resonance energy range. 
The capability of FLUKA for cosmic ray calculations has recently been 
demonstrated \cite{battistoni}. 

The GHEISHA \cite{gheisha} program 
successfully used in the detector Monte Carlo code GEANT3
\cite{geant} is called GHEISHA 600
within this study  to distinguish it from
version GHEISHA 2002 with modified kinematics using correction patches
\cite{gheishafix} which improve energy and momentum conservation.

The {\bf H}illas {\bf S}plitting {\bf A}lgorithm (HSA) \cite{hillas} is 
employed in the EAS simulation code AIRES \cite{aires,sciutto}. 
Its parametrizations are valid only for collisions with air which 
do not allow a direct comparison with experiments performed with other targets.
The HSA has been linked with CORSIKA for test purposes 
only to add the results of these tests to the present comparison. 
It is not planned to make the HSA available within CORSIKA.

The {\bf U}ltra-{\bf r}elativistic {\bf Q}uantum 
{\bf M}olecular {\bf D}ynam\-ic (UrQMD$~$1.3) model \cite{urqmd} 
describes microscopically the projectile transport through 
the air target in tiny steps ($\approx$ 0.2 fm) and follows 
collisions and/or scatterings on the hadron level. 
Nuclear resonance effects are considered in detail.

\section{SINGLE COLLISIONS}
\label{sec-sing_coll}

All simulations presented in this section have been performed with 
the so-called `interactiontest' option of CORSIKA to compare 
all interaction programs under equal conditions.

\subsection{pp-collisions}
\label{sec-pp}

\begin{figure}[t]
 \begin{center}
  \hspace*{-5pt}
  \includegraphics[width=17.8pc]{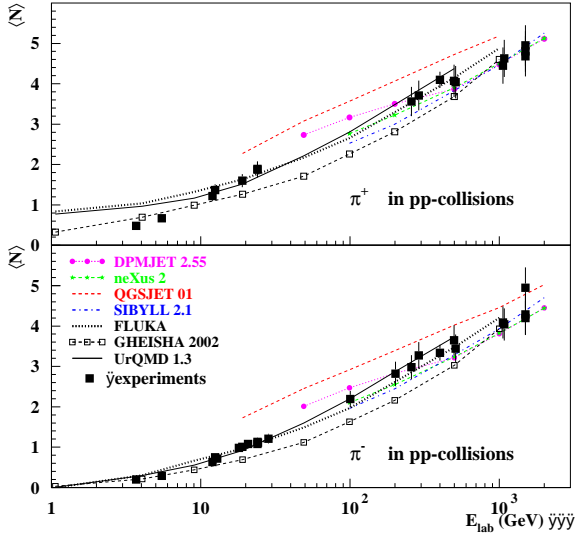}
 \end{center}
 \vspace*{-28pt}
 \caption{$\pi^+$ (upper panel) and $\pi^-$ (lower panel) multiplicities 
          in pp-collisions as function of projectile energy. 
          Experimental points are taken from \cite{antinucci,rossi,tan}.}
 \label{fig-pp_multi}
 \vspace*{-6pt}
\end{figure}

The $\pi^+$  and $\pi^-$-multiplicities in pp-collisions are 
shown in Fig.~\ref{fig-pp_multi}. 
Most of the displayed low-energy models follow the experimental 
points sufficiently well except GHEISHA which produces too few pions.
QGSJET overestimates the pion multiplicity within the 
20 - 500 GeV region at the lower end of its design energy range.

\subsection{p-$^9$Be collisions}
\label{sec-p9Be}

\begin{figure}[t]
  \begin{center}
   \hspace*{-5pt}
   \includegraphics[width=17.8pc]{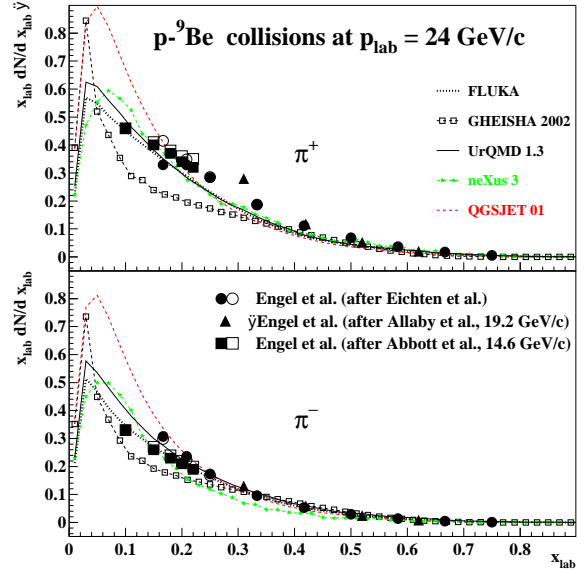}
  \end{center}
  \vspace*{-28pt}
  \caption{ Distributions of $\pi^+$ (upper panel) and $\pi^-$ (lower panel) 
            longitudinal momentum fractions $x_{\rm lab} = p_{\rm tot}/p_{\rm beam}$ 
            in p-$^9$Be collisions at 24 GeV/c. 
            Experimental data were derived by Engel et al. \cite{engel} from 
            \cite{abbott,allaby,eichten}.}
  \label{fig-pBe_plab_24GeV_pip+pim}
  \vspace*{-6pt}
 \end{figure}

The above mentioned deficit of pions produced by GHEISHA is visible 
in the longitudinal momentum fractions of p-$^9$Be collisions 
(Fig.~\ref{fig-pBe_plab_24GeV_pip+pim}) at $x_{\rm lab}$-values 
around 0.15.
As in pp-collisions QGSJET shows a pion overestimation visible 
in Fig.~\ref{fig-pBe_plab_24GeV_pip+pim} at $x_{\rm lab}<0.1$. 
Despite the completely different approaches of FLUKA and UrQMD 
both models reproduce the experimental data equally well. 

For kaons all codes behave similar as for the pions.
The dominance of K$^+$-mesons over the K$^-$  ones caused 
by the associated production of K$^+$+$\Lambda$ 
in proton initiated interactions is reproduced by all studied
models similarly (shown in Ref.~\cite{heck_tsukuba}).
FLUKA produces a small deficit of K$^{\pm}$ 
at $x_{\rm lab} < 0.3$.

\begin{figure}[t]
  \begin{center}
   \hspace*{-5pt}
   \includegraphics[width=17.8pc]{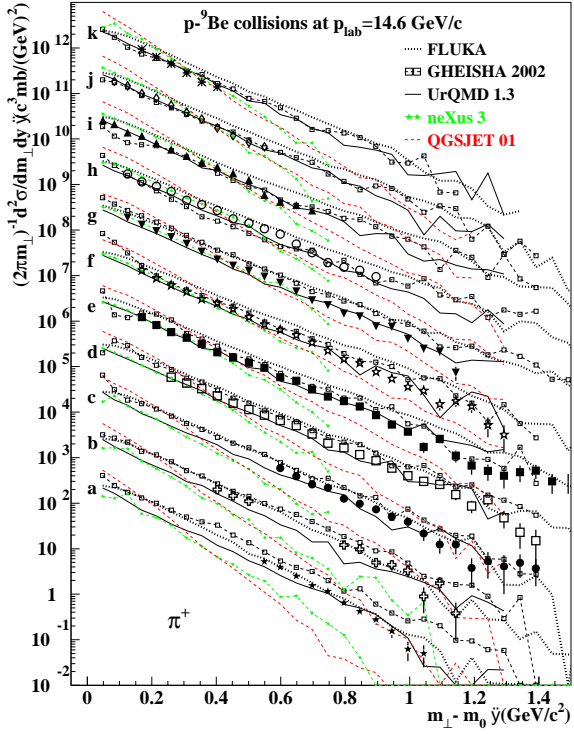}
  \end{center}
  \vspace*{-28pt}
  \caption{Invariant cross sections as function of m$_{\perp}$ for 
           $\pi^{+}$ in p-$^9$Be collisions at 14.6 GeV/c. The curves  
           {\it a} - {\it k} give the distributions for consecutive rapidity 
           intervals $\Delta$y = 0.2 increasing in the range 0.6$<$y$<$2.8. 
           Each curve is multiplied by a factor 10$^{l}$  
           with {\it l} increasing from 0 to 10 to separate the curves.
           Experimental points are taken from \cite{abbott}.}
  \label{fig-pBe_pt_vs_y_14GeV_pip}
  \vspace*{-6pt}
 \end{figure}

The agreement of the transverse momenta with experimental data 
is examined in Fig.~\ref{fig-pBe_pt_vs_y_14GeV_pip} 
where the double-differential cross sections of $\pi^+$-mesons
are given as function of transverse mass and rapidity.
The rapidity bins of width $\Delta \rm y=0.2$ extend from
0.6 (curve {\it a} at bottom) to 2.8 (curve {\it k} on top).
Both FLUKA and UrQMD follow the exponential trend (straight 
lines in logarithmic scale) of the experimental points \cite{abbott} 
with the correct slope in the whole rapidity range. 
GHEISHA exhibits significant deviations for small transverse masses
$\rm m_{\perp}-m_{0}<0.2$ and for rapidities $1.0<{\rm y}<2.4$
(curves {\it c} to {\it i}). 
Presumably the kinematics are still far from being perfect 
despite the correction patches \cite{gheishafix}.
As the high-energy models QGSJET and {\sc neXus} 
use a $p_{\perp}$-parametrization covering a wide
energy range up to highest energies it is not surprising
that below the lowest end of their designed energy range
they produce transverse momenta with slopes generally too steep.
All these findings hold similarly for $\pi^-$-mesons  emerging 
from p-$^9$Be collisions.

\subsection{p-$^{12}$C collisions}
\label{sec-p12C}

\begin{figure}[t]
  \begin{center}
   \hspace*{-5pt}
   \includegraphics[width=17.8pc]{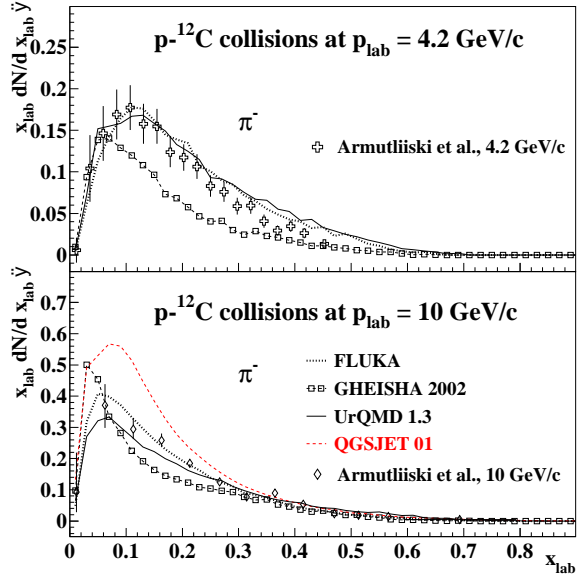}
  \end{center}
  \vspace*{-28pt}
  \caption{ Distribution of $\pi^{-}$ longitudinal momentum fractions
            $x_{\rm lab} = p_{\rm tot}/p_{\rm beam}$ in p-$^{12}$C collisions 
            at 4.2 (upper panel) and 10 GeV/c (lower panel). 
            Experimental data are taken from \cite{armutliiski}.}
  \label{fig-pC_plab_4GeV_10GeV_det}
  \vspace*{-6pt}
 \end{figure}

For the carbon target the diagnoses resemble the
conclusions from Sect.~\ref{sec-p9Be}.
Longitudinal momentum fractions of $\pi^{-}$-mesons are plotted 
for projectile energies of 4.2 and 10 GeV in Fig.~\ref{fig-pC_plab_4GeV_10GeV_det}. 
Again the agreement of FLUKA and UrQMD with the experimental
data is fine. 
Simultaneously for low-energy pions ($x_{\rm lab}~<~0.3$) QGSJET
produces an overshoot and GHEISHA shows a deficit at both 
examined energies. 

\subsection{Collisions with $^{14}$N and air targets}
\label{sec-p14N}

In contrast to the preceding sections no 
experimental data exist for $^{14}$N or air targets. 
We take the behaviour of FLUKA and UrQMD
as reference as both models have shown a sufficiently
good agreement with experimental values for 
lighter targets.

\begin{figure}[t]
  \begin{center}
   \hspace*{-5pt}
   \includegraphics[width=17.8pc]{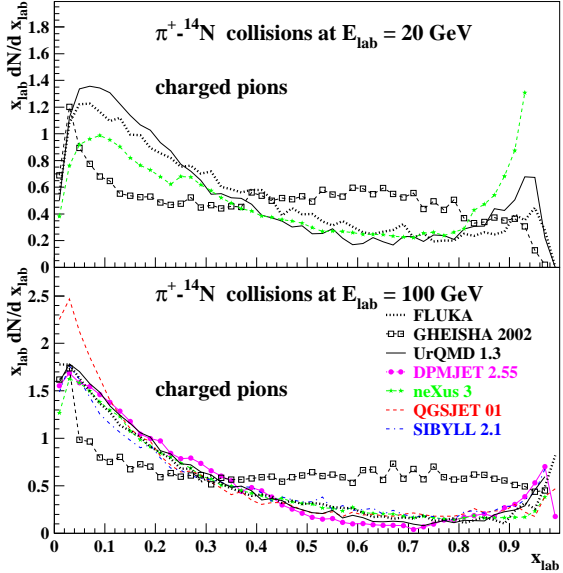}
 \end{center}
  \vspace*{-28pt}
  \caption{ Distribution of $\pi^{\pm}$ longitudinal momentum fractions
            $x_{\rm lab} = p_{\rm tot}/p_{\rm beam}$ in $\pi^+$-$^{14}$N collisions
            at 20 (upper panel) and 100 GeV (lower panel).}
  \label{fig-piNxf_2e10_1e11}
  \vspace*{-6pt}
 \end{figure}

As charged pions are by far the most frequent projectiles 
within an EAS and $^{14}$N is the most abundant target nucleus
within air, we investigate in Fig.~\ref{fig-piNxf_2e10_1e11} 
the momentum fractions which the  $\pi^{\pm}$-mesons  carry away 
from $\pi^+$-$^{14}$N collisions.
The behaviour of the displayed models resembles that 
which has already been found for the lighter targets.
FLUKA and UrQMD are close to each other. 
All other high energy models, 
at 20 GeV collision energy {\sc neXus}, at 100 GeV additionally
DPMJET, QGSJET, and SIBYLL 
result in similar $x_{\rm lab}$-distributions
(Again the overestimation of QGSJET at low $x_{\rm lab}$-values
deviates from the average behaviour).
The peak occurring at $x_{\rm lab} > 0.9$ results
from diffractive interactions.
Differing from all other interaction models
GHEISHA creates the pions with momenta shifted to higher values
for both displayed collision energies, i.e. with an excess  
in the range  $x_{\rm lab} > 0.4$ and a deficit at 
$x_{\rm lab} < 0.3$.
In EAS simulations this finding implies too high an elasticity which
results in a stretching of the low-energy branches in the hadronic backbone.
\vspace{3pt}

In p-air interactions we include the HSA \cite{hillas} as used in AIRES 
2-6-0 \cite{sciutto}.
It is parametrized only for air as target.
\begin{figure}[t]
 \begin{center}
  \hspace*{-5pt}
  \includegraphics[width=17.8pc]{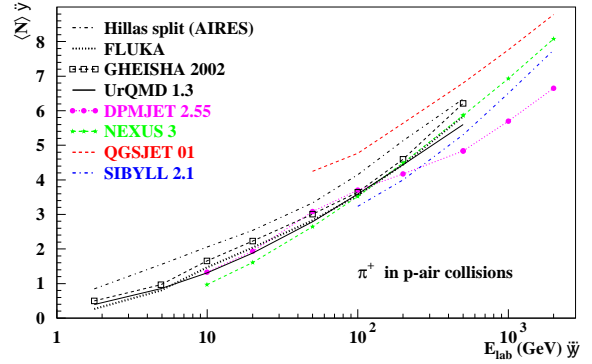}
 \end{center}
 \vspace*{-28pt}
 \caption{Multiplicities of $\pi^+$ in p-air collisions as 
          function of projectile energy.}
 \label{fig-p-air_pip_multi}
 \vspace*{-6pt}
\end{figure}
%
\begin{figure}[h]
 \begin{center}
  \hspace*{-5pt}
  \includegraphics[width=17.8pc]{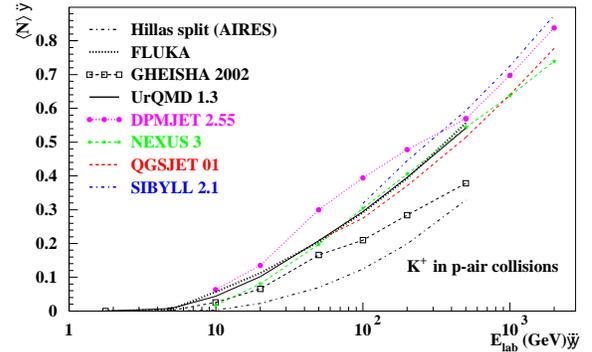}
 \end{center}
 \vspace*{-28pt}
 \caption{Multiplicities of K$^+$ in p-air collisions as 
          function of projectile energy.}
 \label{fig-p-air_kp_multi}
 \vspace*{-6pt}
\end{figure}
Fig.~\ref{fig-p-air_pip_multi} shows the $\pi^+$-multiplic\-ities as function
of the collision energy.
Assuming FLUKA and 
UrQMD as standard we observe an overshoot of QGSJET
as in Sect.~\ref{sec-pp}.
For GHEISHA the multiplicities seem now to be in agreement with the 
reference models.

The HSA as implemented in AIRES has been "{\it configured to approximately emulate the 
multiplicities and energy distributions of other models}" \cite{sciutto}.
An interpolation between GHEISHA and QGSJET has been selected \cite{knapp}
to get a smooth transition to the high-energy model QGSJET. 
\begin{figure}[t]
 \begin{center}
  \hspace*{-5pt}
  \includegraphics[width=17.8pc]{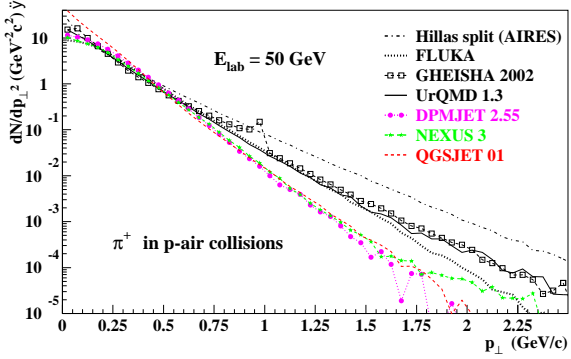}
 \end{center}
 \vspace*{-32pt}
 \caption{Transverse momentum distribution of $\pi^+$ in p-air collisions  
          at 50 GeV.}
 \label{fig-p-air_pt3_pip_50GeV}
 \vspace*{-12pt}
\end{figure}
%
\begin{figure}[h]
 \begin{center}
  \hspace*{-5pt}
  \includegraphics[width=17.8pc]{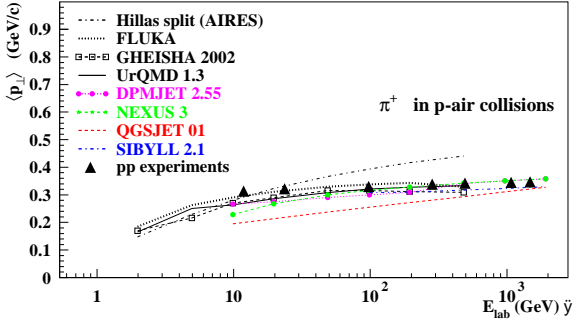}
 \end{center}
 \vspace*{-32pt}
 \caption{Average transverse momentum of $\pi^+$ in p-air collisions as 
          function of projectile energy. Experimental values of pp-collisions
           are taken from \cite{rossi}.}
 \label{fig-p-air_pt3_pip}
 \vspace*{-12pt}
\end{figure}
In my opinion HSA should not be tuned to imitate other models, 
rather the parameters should be adjusted in a manner to approach  
experimental values as closely as possible.

The K$^+$-multiplicities displayed in Fig.~\ref{fig-p-air_kp_multi}
show a nice agreement of most models for energies below $\approx$50 GeV.
In the HSA the K$^+$-number has a significant deficit. 
By a comparison with the corresponding K$^-$-multiplicities the 
discrepancy can be traced back to kaons resulting from the 
associated K$^+$+$\Lambda$ production which is missing in the HSA. 
GHEISHA exhibits a moderate deficit for both charged K-mesons which 
increases with the collision energy.
Generally for negative kaon multiplicities the models show much better 
agreement than for the positive ones.

Special care has to be taken of the transverse momenta as they might influence 
the lateral distribution of the EAS particles arriving at ground.  
In the HSA the transverse momentum distributions need to be inserted 
from outside. 
For the other models we have examined the $p_{\perp}$-feature already 
in p-$^{9}$Be collisions (see Fig.~\ref{fig-pBe_pt_vs_y_14GeV_pip}). 
Fig.~\ref{fig-p-air_pt3_pip_50GeV} compares the $p_{\perp}$-distributions
of $\pi^{+}$-mesons for all low-energy and several high-energy codes. 
The reference models FLUKA and UrQMD  coincide up to $p_{\perp}$$<$ 1.75 GeV/c.
Their predictions are well separated from that of HSA with its
flat slope indicating an insufficient approximation to
experimental values. 
This flat slope is detectable in the whole design range of the
HSA within AIRES \cite{aires}.
The GHEISHA distribution exhibits even a completely unexpected peak at 
$\approx$$1$~GeV/c.  
As discussed in Sect.~\ref{sec-p9Be}, with increasing $p_{\perp}$
the high-energy models fall-off with a too steep slope.

\begin{figure}[t]
 \begin{center}
  \hspace*{-5pt}
  \includegraphics[width=17.8pc]{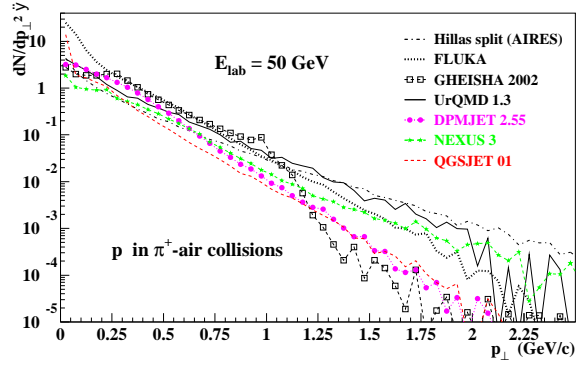}
 \end{center}
 \vspace*{-32pt}
 \caption{Transverse momentum distribution of protons in $\pi^+$-air 
          collisions at 50 GeV.}
 \label{fig-pi-air_pt_prot_50GeV}
 \vspace*{-12pt}
\end{figure}

The average transverse momentum $\langle p_{\perp} \rangle$ as function of 
collision energy is given in Fig.~\ref{fig-p-air_pt3_pip}. 
Nearly all models follow the trend of experimental values
of pp-collisions. 
HSA shows a too strong increase of $\langle p_{\perp} \rangle$ with 
collision energy which results from the flat transverse momentum 
distribution  shown in Fig.~\ref{fig-p-air_pt3_pip_50GeV}.
The parametrization of QGSJET below the lower end of its 
designed energy range results in too low an
average transverse momentum, especially below $E_{\rm lab} < 500$ GeV.

For pion-induced collisions with air 
the transverse momentum distributions
of protons are displayed in Fig.~\ref{fig-pi-air_pt_prot_50GeV}.
Again the HSA shows a deficit at small transverse momenta and an overshoot
at large $p_{\perp}$-values. 
The kink in the $p_{\perp}$-distribution of GHEISHA at 
$\approx 1$ GeV/c once more demonstrates the problem in the
kinematics of this model.

\section{AIR SHOWER SIMULATIONS}
\label{sec-shower_sim}

The low-energy hadronic interaction codes have been combined with 
the high-energy code QGSJET$~$01 and in part with SIBYLL$~$2.1
to simulate p-induced showers of 10$^{19}$ eV
at vertical incidence.
In the simulations the thinning method \cite{hillas,kobal} is employed
with the parameters given in Table \ref{tab-cpu}.
For UrQMD the thinning level has been modified to 
$\varepsilon$=$10^{-5}$ because of the long required CPU-times.
For all combinations the transition energy between high
and low-energy model has been taken as 80 GeV.
100 showers of each model combination have been averaged to 
minimize the influence of shower-to-shower fluctuations.

\subsection{Longitudinal profiles}
\label{sec-longi}

\begin{figure}[t]
 \begin{center}
  \hspace*{-5pt}
  \includegraphics[width=17.8pc]{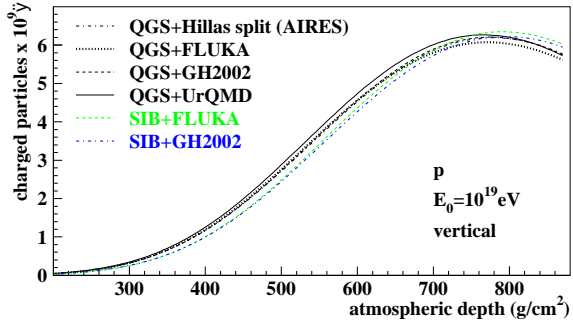}
 \end{center}
 \vspace*{-32pt}
 \caption{Longitudinal profiles of charged particle numbers
          for different combinations of interaction models.}
 \label{fig-long_1e19}
 \vspace*{-12pt}
\end{figure}

The resulting longitudinal profiles are reproduced in 
Fig.~\ref{fig-long_1e19}.  
All averaged profiles with QGSJET 
reach the shower maximum at $777 \pm 10~$g/cm$^2$. 
The differences exceed slightly the standard deviation of 
6.5 g/cm$^2$.  UrQMD with enhanced thinning fluctuations reveals 
the smallest value at 766 g/cm$^2$, while FLUKA penetrates deepest
($X_{\rm max} = 786~$g/cm$^2$). 
For the combinations with SIBYLL the value       
$X_{\rm max}~=~791~$g/cm$^2$ of FLUKA 
agrees well with 796$~$g/cm$^2$ of GHEISHA 
(for the influence of high-energy hadronic interaction models on 
$X_{\rm max}$ see Fig.~6 in Ref.~\cite{heck_cern}).

One might argue the differences coming from different production 
cross sections, but for nucleon and pion projectiles with 
$p_{\rm lab} > 1~$GeV/c the corresponding cross sections agree within 
$\pm 5$ \%, only the kaon cross sections show somewhat larger differences.
No correlation of the $X_{\rm max}$-values with 
the cross sections of the low-energy models could be observed.

\subsection{Lateral distributions}
\label{sec-late}

The lateral particle number distributions fall off 
by 5 orders of magnitude in the distance range from 200 m to 5 km.
\begin{figure}[t]
 \begin{center}
  \hspace*{-5pt}
  \includegraphics[width=17.8pc]{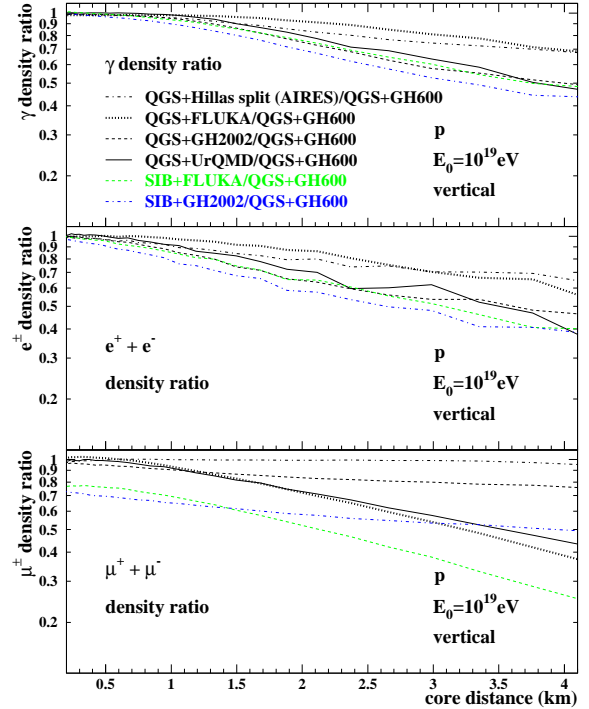}
 \end{center}
 \vspace*{-32pt}
 \caption{Lateral distribution ratios of particle numbers 
          for $\gamma$ (top), e$^{\pm}$ (middle), and $\mu^{\pm}$ (bottom)
          relative to the model combination QGSJET$~$01 + GHEISHA$~$600.}
 \label{fig-lat_1e19_ratios}
 \vspace*{-12pt}
\end{figure}
Therefore the ratios relative to the combination QGSJET + GHEISHA$~$600 
are displayed in Fig.~\ref{fig-lat_1e19_ratios}.
This combination  of high and low-energy models shows the flattest 
distribution for $\gamma$, e$^{\pm}$, and $\mu^{\pm}$-densities.
The reason might be the wrong kinematics resulting in a prolongated 
shower development at the low-energy end of the hadronic backbone.
Close to the shower axis this effect is masked by the numerous particles 
emerging from high-energy interactions, while at large core distances 
with a longer shower development path and therefore larger hadronic 
shower age the excess in elasticity of GHEISHA 600 produces more 
secondaries which finally flattens the lateral distributions. 
Similarly the high elasticity of GHEISHA 2002 found in 
Sect.~\ref{sec-p14N} for $\pi^+$-$^{14}$N collisions
may explain the flatter distributions (with QGSJET and SIBYLL) 
for the $\mu^{\pm}$-density.
\begin{figure}[t]
 \begin{center}
  \hspace*{-5pt}
  \includegraphics[width=17.8pc]{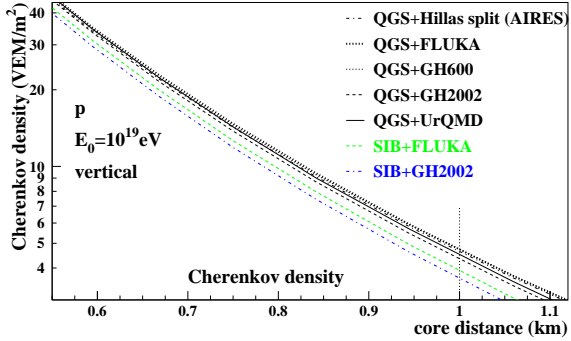}
 \end{center}
 \vspace*{-32pt}
 \caption{Lateral distributions of Cherenkov densities.}
 \label{fig-lat_cher_dens}
 \vspace*{-12pt}
\end{figure}
%
\begin{figure}[t]
 \begin{center}
  \hspace*{-5pt}
  \includegraphics[width=17.8pc]{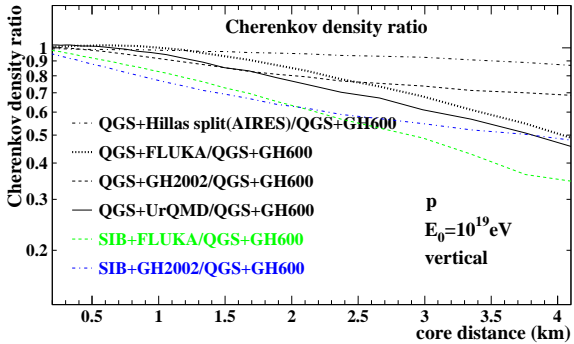}
 \end{center}
 \vspace*{-32pt}
 \caption{Lateral distribution ratios of Cherenkov densities relative to
          the model combination QGSJET$~$01 + GHEISHA$~$600.}
 \label{fig-lat_cher_dens_ratio}
 \vspace*{-12pt}
\end{figure}

While FLUKA and UrQMD show similar distributions for 
the $\mu^{\pm}$-densities (and to a minor degree for the  
e$^{\pm}$-densities), the flat distributions of HSA have to be 
attributed to the insufficient $p_{\perp}$-parametrization.
The combinations with SIBYLL exhibit a significantly lower 
$\mu^{\pm}$-density even at core distances as small as 1 km. 

For the energy calibration of the Auger experiment \cite{auger} the signal 
S(1000) of the Cherenkov water tanks at 1000 m core distance will be used. 
Here we investigate the dependence of S(1000) on the low-energy model.
With the assumptions for the Cherenkov water tanks that \\[-20pt]
\begin{itemize}
\item em-particles deposit their full energy,\\[-20pt]
\item muons contribute with their full energy, 
      but at maximum with 240 MeV, and \\[-20pt]
\item 240 MeV deposited energy corresponds to 1 vertical 
      equivalent muon (VEM) \\[-20pt]
\end{itemize} 
the lateral Cherenkov densities given in Fig.~\ref{fig-lat_cher_dens} 
are derived. 
For the Auger tanks of 10~m$^2$ area 
the signal S(1000) of a vertical 10$^{19}~$eV shower (p-induced) amounts
to 47 (QGSJET+FLUKA), 46 (QGSJET+HSA), 45 (QGSJET+UrQMD), 
43 (QGSJET+GHEISHA 2002), 39 (SIBYLL+FLUKA), and 
36 (SIBYLL+ GHEISHA 2002) VEM.
Radial distributions of Cherenkov densities are plotted 
in Fig.~\ref{fig-lat_cher_dens_ratio} as ratios which again display 
the large dependence on the high-energy model at 1 km core distance.
The differences increase at larger distances still relevant 
to measurements of the Auger detector.

\section{IMPLICATIONS AND CONCLUDING REMARKS}
\label{sec-implicat}

The CPU-time requirements to treat single collisions rsp.~complete 
showers is important for EAS-simulations.
In Table \ref{tab-cpu} CPU-times are collected.

The HSA-code as the fastest has an unsatisfactory performance as 
demonstrated in the preceding sections and lacks predictive power. 
A revision of its parameters is needed.

As shown in previous sections the kinematics of GHEISHA 2002 is still
not satisfying. 
\begin{table}[b]
\vspace{-12pt}
\begin{center}
\caption{CPU times for DEC-alpha 1000XP (500 MHz).}
\label{tab-cpu}
\begin{tabular}{l|r|r|r}
\hline
        &\multicolumn{2}{c|}{$10^5$ collisions}&
                                   \multicolumn{1}{c}{1 shower $^1$} \\[-1pt]
        & \multicolumn{2}{c|}{at 10 GeV}&
                                   \multicolumn{1}{c}{$10^{19}$ eV}\\[-1pt]
$~~~$Model& p-air        & $\pi$-air &
                              \multicolumn{1}{c}{$\varepsilon=10^{-6}$}\\
\hline
FLUKA        &        181   &   164           & 63300 \\
GHEISHA$~$2002&       108   &   102           & 29100 \\
HSA (AIRES)  &         64   &    64           & 18000 \\  
UrQMD 1.3 $^2$ &    12200   & 11400           &($\approx$800000)\\
\hline
{\sc neXus 3}&       6173   &  5861           &           \\
QGSJET$~$01  &         88   &    87           &            \\
\hline
\multicolumn{4}{l}{\scriptsize $^1$ QGSJET 01, $\theta = 0^\circ$, 
                 $E_{\rm h}>300$MeV, $E_{\rm \mu}>100$MeV,} \\[-2pt]
\multicolumn{4}{l}{\scriptsize $~~$ $E_{\rm em}>250$keV, 
                 W$_{\rm em}<10^4$, W$_{\rm h}<10^2$.}\\[-1pt]
\multicolumn{4}{l}{\scriptsize $^2$ H.J. Drescher has accelerated UrQMD 1.3c by 
                                                     a factor 15.} \\[-2pt]
\end{tabular}
\end{center}
 \vspace*{-21pt}
\end{table}

The microscopic UrQMD model gives reliable results 
at the expense of extreme long CPU-times. 
In this model no parametrized cross sections are used, rather the
projectile hits on a `disk' chosen large enough to cover the 
maximum collision parameter.
Several trials (with their time-consuming microscopic calculation) 
without interaction are  usually needed before in an 
inelastic collision secondary particles are produced. 
Therefore this model is less suited for EAS simulations.\\[3pt]
The FLUKA model shows the best performance and it is recommended 
for EAS-simulations despite the longer CPU-times.
Unfortunately its source code is not yet publicly available, 
only object codes for 5 types of machine/operating 
systems are distributed at present.

Finalizing one can state that measurable shower parameters 
are predominantly influenced by the high-energy interaction programs
\cite{heck_cern} and only to some minor extend by the low-energy models.

\section*{Acknowledgments}
I am indebted to the authors of the interaction models 
for their advice in linking their codes with CORSIKA. 
Special thanks go to  G. Battistoni and  A. Ferrari 
who enabled the use of FLUKA with CORSIKA. 
The support of M. Bleicher, H.J. Drescher, S. Soff, H. St{\"o}cker, and 
H. Weber to get the UrQMD 1.3 running for EAS simulations is acknowledged.  
I thank R. Engel who provided me with experimental 
data sets and helped with clarifying discussions.
Last but not least my thanks go to K. Bekk and J. Manger for their 
continuous efforts to keep the DEC-workstations running which was 
a prerequisite for this study.


\end{document}